\PassOptionsToPackage{pdfpagelabels=false}{hyperref}

\documentclass[fleqn,usenatbib,useAMS]{mnras}

\usepackage{graphicx}
\usepackage{amsmath}
\usepackage[dvipsnames]{xcolor}  
\usepackage{amssymb}
\usepackage{color}
\usepackage{ulem}
\usepackage{soul}
\usepackage[figure,figure*,table,table*]{hypcap}
\usepackage{subfigure}
\usepackage{multirow}

\title[Mass and scale-dependence of lensing is low]{On the halo-mass and radial scale dependence of the lensing is low effect}
\author[J. U. Lange et al.]{Johannes U. Lange$^{1, 2}$\thanks{email: jolange@ucsc.edu}, Alexie Leauthaud$^1$, Sukhdeep Singh$^{3, 4}$, Hong Guo$^5$, \newauthor Rongpu Zhou$^6$, Tristan L. Smith$^7$ and Francis-Yan Cyr-Racine$^8$\\
$^1$Department of Astronomy and Astrophysics, University of California, Santa Cruz, CA 95064, USA\\
$^2$Kavli Institute for Particle Astrophysics and Cosmology and Department of Physics, Stanford University, CA 94305, USA\\
$^3$Berkeley Center for Cosmological Physics, University of California, Berkeley, CA 94720, USA\\
$^4$McWilliams Center for Cosmology, Department of Physics, Carnegie Mellon University, Pittsburgh, PA 15213, USA\\
$^5$Key Laboratory for Research in Galaxies and Cosmology, Shanghai Astronomical Observatory, Shanghai 200030, China\\
$^6$Lawrence Berkeley National Laboratory, Berkeley, CA 94720, USA\\
$^7$Department of Physics and Astronomy, Swarthmore College, Swarthmore, PA 19081, USA\\
$^8$Department of Physics and Astronomy, University of New Mexico, Albuquerque, NM 87106, USA}

\pubyear{2020}

\begin{document}

\date{Accepted xxx. Received xxx}

\label{firstpage}
\pagerange{\pageref{firstpage}--\pageref{lastpage}}

\maketitle

\begin{abstract}
    The canonical $\Lambda$CDM cosmological model makes precise predictions for the clustering and lensing properties of galaxies. It has been shown that the lensing amplitude of galaxies in the \textit{Baryon Oscillation Spectroscopic Survey} (BOSS) is lower than expected given their clustering properties. We present new measurements and modelling of galaxies in the BOSS LOWZ sample. We focus on the radial and stellar mass dependence of the lensing amplitude mis-match. We find an amplitude mis-match of around $35\%$ when assuming $\Lambda$CDM with Planck Cosmological Microwave Background (CMB) constraints. This offset is independent of halo mass and radial scale in the range $M_{\rm halo}\sim 10^{13.3} - 10^{13.9} h^{-1} M_\odot$ and  $r=0.1 - 60 \, h^{-1} \mathrm{Mpc}$ ($k \approx 0.05 - 20 \, h \, {\rm Mpc}^{-1}$). The observation that the offset is both mass and scale independent places important constraints on the degree to which astrophysical processes (baryonic effects, assembly bias) can fully explain the effect. This scale independence also suggests that the ``lensing is low" effect on small and large radial scales probably have the same physical origin. Resolutions based on new physics require a nearly uniform suppression, relative to $\Lambda$CDM predictions, of the amplitude of matter fluctuations on these scales. The possible causes of this are tightly constrained by measurements of the CMB and of the low-redshift expansion history.  
\end{abstract}


\section{Introduction}

The $\Lambda$ cold dark matter ($\Lambda$CDM) model makes precise predictions about the large-scale structure properties of the Universe. In this model, the expansion history of the Universe is determined by radiation, matter, and dark energy ($\Lambda$); and the growth of structure follows that of a collisionless fluid called dark matter. Large galaxy surveys map the matter field via galaxies that reside in gravitationally collapsed structures called dark matter haloes. Given the wealth of information available from current observations, one can make testable predictions via the $\Lambda$CDM model despite not knowing a priori how galaxies occupy dark matter haloes, a relationship called the galaxy-halo connection.

In recent years, fueled by the increasing precision of cosmological measurements, there is mounting evidence that the canonical $\Lambda$CDM fails at correctly predicting observations. At the forefront of this tension between $\Lambda$CDM and observations are comparisons between inferences from the cosmic microwave background (CMB) and low-redshift observations of the nearby Universe. The most significant finding is the so-called $H_0$-tension: observations of the CMB \citep{PlanckCollaboration2020_AA_641_6} infer a lower value for the present-day expansion rate of the Universe than direct measurements \citep[see e.g.][]{Riess2019_ApJ_876_85}. Ultimately, this finding could point to revisions to our standard $\Lambda$CDM model \citep[see][and references therein]{Knox2020_PhRvD_101_3533}.

In addition to the $H_0$ tension, there is also increasing evidence that CMB predictions for the amount of structure in the low-redshift Universe do not match with observations. The tension is commonly expressed in terms of constraints on $S_8 = \sigma_8 \sqrt{\Omega_{\rm m} / 0.3}$ where $\sigma_8$ is strength of matter fluctuations and $\Omega_{\rm m}$ is the fraction of the matter-energy density of the Universe in matter. For example, a variety of studies analysing galaxy clustering and galaxy-galaxy lensing favour values for $S_8$ that are around $\sim 15\%$ lower than the values preferred by the \cite{PlanckCollaboration2020_AA_641_6} analysis \citep{Cacciato2013_MNRAS_430_767, Leauthaud2017_MNRAS_467_3024, Abbott2018_PhRvD_98_3526, Lange2019_MNRAS_488_5771, Yuan2020_MNRAS_493_5551, Singh2020_MNRAS_491_51}. The statistical significance of the discrepancy for each of these low-redshift studies and the CMB is at the level of $\gtrsim 2\sigma$ depending on the data and scales analysed. Similar tensions have been found through studies of clusters selected by the Sunyaev-Zeldovich effect \citep{PlanckCollaboration2016_AA_594_24}, cosmic shear \citep{Troxel2018_PhRvD_98_3528, Hikage2019_PASJ_71_43, Hildebrandt2020_AA_633_69} and the Lyman-$\alpha$ power spectrum \citep{PalanqueDelabrouille2020_JCAP_04_038}.

An alternative manifestation of this problem is that models for the galaxy-halo connection fit to the clustering properties of galaxies do not correctly predict their galaxy-galaxy lensing amplitudes if the best-fit cosmological parameters of the \cite{PlanckCollaboration2020_AA_641_6} analysis are assumed. \cite{Leauthaud2017_MNRAS_467_3024} show that different models for the galaxy-halo connection in the \textit{Baryon Accoustic Oscillation Survey} (BOSS) CMASS sample overpredict the measured lensing signal by around $40\%$ on non-linear scales. This finding was later also confirmed for the BOSS LOWZ sample \citep{Lange2019_MNRAS_488_5771, Singh2020_MNRAS_491_51, Wibking2020_MNRAS_492_2872}. Additionally, it was shown that assembly bias, the often neglected effect that the clustering amplitudes of dark matter halos depend on halo properties besides mass, cannot fully account for the mismatch in the lensing amplitudes \citep[][]{Lange2019_MNRAS_488_5771, Yuan2020_MNRAS_493_5551}. Similarly, it was shown that the impact of baryons on the matter distribution on small scales is insufficient to explain the observations, both based on predictions from hydrodynamical simulations \citep{Leauthaud2017_MNRAS_467_3024, Lange2019_MNRAS_488_5771} and via constraints from observations of the thermal Sunyaev-Zeldovich effect \citep{Amodeo2020_arXiv_2009_5558}. Recently, \cite{Zu2020_arXiv_2010_1143} asked whether or not extreme galaxy-halo models could explain the lensing-is-low effect on small scales. However, we argue later that this would require satellite fractions that are likely to be inconsistent with other observations.

Interestingly, a recent study of the \textit{Dark Energy Survey} (DES) found evidence that the relative amount by which the lensing amplitude is over-predicted could depend on host halo mass \citep{Abbott2020_PhRvD_102_3509}. They show that the cosmological constraints they obtain from a combination of cluster abundance and weak lensing masses depends on the cluster sample analysed. Particularly, clusters with low richness, i.e. few satellite galaxies, prefer lower values for $S_8$. Similarly, the relative over-prediction of the lensing amplitudes was shown to be the strongest for galaxies living in low-mass haloes. If not due to observational systematics \citep{Abbott2020_PhRvD_102_3509}, this finding would place interesting constraints on theoretical models explaining the lensing over-prediction. For example, changes in the cosmological parameter $S_8$ would have a roughly mass-independent impact on the predicted lensing signal at fixed clustering \citep{Lange2019_MNRAS_488_5771}. Similarly, for the hydrodynamical simulations analysed in \cite{Lange2019_MNRAS_488_5771}, there was also no strong halo mass dependence to the relative impact of the galaxy-galaxy lensing amplitude.

The goal of the present work is to analyse the mass and radial dependence of the mismatch between predicted and observed lensing amplitude under the \cite{PlanckCollaboration2020_AA_641_6} $\Lambda$CDM cosmology. In this work, we do not explicitly model the effects of galaxy assembly bias and baryonic feedback. These two effects have already been studied elsewhere \citep[][]{Leauthaud2017_MNRAS_467_3024,Lange2019_MNRAS_487_3112,Yuan2020_MNRAS_493_5551} and both effects have been shown to be important on smaller scales, $r \lesssim 5 \, h^{-1} \mathrm{Mpc}$, but are complex and non-trivial to model. Instead, our findings on the mass and scale dependence places model-independent constraints on these and physical explanations for the lensing amplitude mismatch. To this end, we analyse the clustering and lensing properties of galaxies in the BOSS LOWZ galaxy sample. Specifically, we analyse LOWZ galaxy samples selected by stellar mass which is known to be correlated with halo mass. Additionally, by analyzing different stellar mass estimates, we can also place limits on which mass estimates correlate more strongly with halo mass \citep{Tinker2017_ApJ_839_121}. Because of the tight correlation between stellar and halo mass \citep{Wechsler2018_ARAA_56_435}, a strong correlation between a stellar mass estimate and halo mass could be seen as indication for a stellar mass estimate being more accurate, i.e. more strongly correlated with the intrinsic stellar mass.

This work extends \citet[][]{Leauthaud2017_MNRAS_467_3024} and \citet{Lange2019_MNRAS_487_3112} to larger radial scales and presents higher signal-to-noise lensing measurements. This work extends \citet[][]{Singh2020_MNRAS_491_51} to smaller radial scales and adds in new constraints on the halo mass dependence of the ``Lensing is Low" effect.

Throughout this work, we assume a $\Lambda$CDM cosmology with $\Omega_{{\rm m}, 0} = 0.307$ for our clustering and lensing measurements. All scales and lensing amplitudes reported are in comoving units and scaled by $h = H_0 / (100 \, \mathrm{km} \, \mathrm{s}^{-1} \, \mathrm{Mpc}^{-1})$ to be independent of the choice of $h$. Our paper is structured as follows. In section \ref{sec:observations}, we describe our observational data and measurements. The modelling framework is described in section \ref{sec:modeling}. We present our main results in section \ref{sec:results} and discuss them in section \ref{sec:discussion}. Finally, our conclusions are presented in section \ref{sec:conclusion}.

\section{Observations}
\label{sec:observations}

Our sample of galaxies is drawn from the BOSS DR12 LOWZ large-scale structure sample \citep{Reid2016_MNRAS_455_1553}. The BOSS LOWZ selection primarily targets galaxies in the redshift range $0.1 \lesssim z \lesssim 0.45$. For our analysis, we only study galaxies in the narrower redshift range $0.2 \leq z \leq 0.35$ to avoid having to model redshift evolution effects in the survey. As discussed in \cite{Ross2017_MNRAS_464_1168}, galaxies in the North (NGC) and the South Galactic Cap (SGC) regions of BOSS have slightly different photometry and thereby target selections. To avoid systematic errors, we only consider galaxies from the larger NGC area.

\subsection{Stellar Masses}

We analyse three different stellar mass, $M_\star$, estimates. The first two are directly derived from SDSS data: the ``Wisconsin'' masses based on a Principal Component Analysis (PCA) of the BOSS spectra \citep{Chen2012_MNRAS_421_314} and the ``Granada'' stellar masses\footnote{For the Granada masses, we use the median posterior mass of the publicly available data, not the best-fit mass. We find that the former correlates more strongly with the other two stellar mass estimates and also results in a stronger clustering of the most massive galaxies, indicating a stronger correlation with halo mass \citep{Tinker2017_ApJ_839_121}.} based on photometry \citep{Ahn2014_ApJS_211_17}. For the ``Granada'' estimates, we utilise results assuming a wide prior on the star-formation history and the possibility for dust extinction. In both cases, we use the results from a \cite{Kroupa2002_Sci_295_82} initial mass function (IMF). 

Finally, we use a new stellar mass estimate based on deeper photometry from the DESI Legacy Imaging Surveys DR8 \citep{Dey2019_AJ_157_168, Zhou2021_MNRAS_501_3309}. First, objects in the Legacy surveys have been cross-matched with SDSS spectroscopic targets, including BOSS LOWZ. This allows us to access deeper photometric data, including near-IR bands from the Wide-field Infrared Survey Explorer \citep[WISE; ][]{Lang2014_AJ_147_108}. To convert photometry to stellar mass estimates, Legacy targets are cross-correlated with galaxies from the Stripe 82 Massive Galaxy Catalog \citep[MGC; ][]{Bundy2015_ApJS_221_15} that have spectroscopic redshifts. Then, a random forest is trained to reproduce near-IR masses from the MGC given the Legacy photometry. This can be done with a precision of around $0.1 \, \mathrm{dex}$ and no strong systematic shift. The trained random forest is then applied to all BOSS LOWZ targets and their associated Legacy fluxes. We note that for around $5\%$ of all BOSS LOWZ targets, no Legacy photometry can be associated. In this case, we use a Wisconsin stellar mass as a proxy: If a galaxy originally without a Legacy stellar mass estimate has a Wisconsin mass estimate placing it into the $n^{\rm th}$ percentile of all Wisconsin mass estimates, we assign it the $n^{\rm th}$ percentile of all Legacy masses.

In order to study the mass-dependence of the lensing and clustering properties, we bin galaxies into three bins according to their stellar mass estimates. The bin edges are defined by $[11.3, 11.5, 11.7, \infty]$, $[11.4, 11.57, 11.75, \infty]$, $[11.1, 11.3, 11.5, \infty]$ in $\log M_\star / M_\odot$ for the Wisconsin, Granada and Legacy mass estimates, respectively. In all cases, the lower bin edges roughly mark the top $95^{\rm th}$, $50^{\rm th}$ and $12^{\rm th}$ percentiles of all masses. In general, we expect higher stellar masses to correlate with higher clustering and lensing amplitudes. The amount of correlation is related to how well the stellar mass estimates trace the host dark matter halo mass \citep{Tinker2017_ApJ_839_121}. We use the observed stellar mass function (SMF) as constraint on our galaxy-halo connection models. We use five mass bins starting from $\log M_\star = 11.3$, $11.4$ and $11.1$ for the Wisconsin, Granada and Legacy masses, respectively. The first four bins have widths of $0.1 \, \mathrm{dex}$ whereas the last bin goes to $\log M_\star = \infty$. We assume a constant, uncorrelated $5\%$ observational uncertainty for all SMF bins when fitting the data. This uncertainty, which is larger than the actual observational uncertainty, is chosen to to not let small details of the SMF strongly affect fits on the galaxy-halo connection.

\subsection{Clustering}

We estimate galaxy clustering using the projected correlation function, $w_{\rm p}$,
\begin{equation}
    w_{\rm p} (r_{\rm p}) = \int\limits_{-r_{\pi, \rm max}}^{+r_{\pi, \rm max}} \xi_{\rm gg} (r_\pi, r_{\rm p}) d r_\pi \, ,
\end{equation}
where $\xi_{\rm gg}$ is the 3D galaxy two-point correlation function and $r_{\rm p}$ and $r_\pi$ are the projected and perpendicular coordinates, respectively. As the integration boundary we choose $\pi_{\rm max} = 100 \, h^{-1} \, \mathrm{Mpc}$. We measure $w_{\rm p}$ in $14$ logarithmic bins in $r_{\rm p}$ going from $0.1 \, h^{-1} \mathrm{Mpc}$ to $63 \, h^{-1} \mathrm{Mpc}$. The two-point correlation function $\xi_{\rm gg}$ is estimated with the \cite{Landy1993_ApJ_412_64} estimator. Additionally, we use the algorithm developed in \cite{Guo2012_ApJ_756_127} to correct for the impact of spectroscopic incompleteness due to fibre collisions.

\begin{figure}
    \centering
    \includegraphics{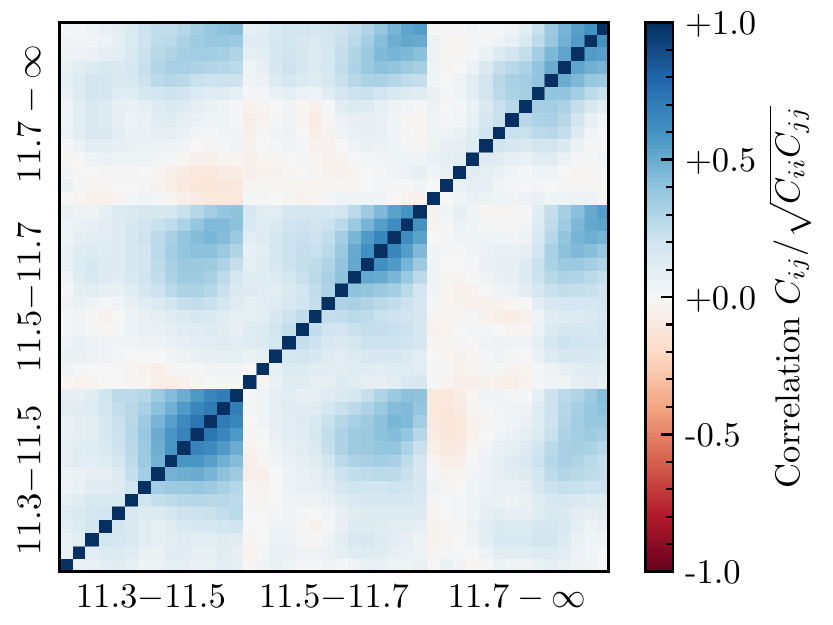}
    \caption{Assumed correlation matrix for the $w_{\rm p}$ measurements. The labels indicate the stellar mass bins in $\log M_\star$. The projected radius $r_{\rm p}$ increases from left to right and bottom to top. The results here are for the Wisconsin stellar masses. The correlation matrices for the Granada and Legacy mass estimates are qualitatively and quantitatively similar.}
    \label{fig:covariance}
\end{figure}

Uncertainties on the measurements are estimated from jackknife re-sampling of $75$ roughly equal size areas. Because of the non-negligible noise in the covariance matrix estimate, we apply a Gaussian smoothing with a scale of $1$ bin for bins close in $r_{\rm p}$ to the correlation matrix. We neglect the diagonal terms of the correlation matrix which are unity by definition. See \cite{Mandelbaum2013_MNRAS_432_1544} for a similar approach. We show the resulting correlation matrix $C_{w_{\rm p}}$ in Fig.~\ref{fig:covariance}. We see non-negligible correlations, especially at large $r_{\rm p}$, even between different stellar mass bins.

\subsection{Lensing}

We estimate the so-called excess surface density, $\Delta\Sigma$,
\begin{equation}
    \Delta\Sigma(r) = \langle \Sigma(<r) \rangle - \Sigma(r),
\end{equation}
by cross-correlating BOSS galaxies with background galaxy shape measurements from SDSS. We follow the same methodology as in \cite{Singh2020_MNRAS_491_51}. Specifically, we use the shape catalogue presented in \cite{Reyes2012_MNRAS_425_2610}. Our estimator for the excess surface density is
\begin{equation}
    \Delta\Sigma = f_{\rm bias} \frac{1 + m}{2 \mathcal{R}} (\Delta\Sigma_{\rm L} - \Delta\Sigma_{\rm R}).
\end{equation}
Here, $f_{\rm bias} = 1.1$ is a correction factor for photometric redshift errors, i.e. it corrects for biases due to photometric redshift inaccuracy and sources physically in front of the lenses. Furthermore, $1 + m = 1.04$ is a correction for shear biases and $\mathcal{R} = 0.87$ the shear responsivity correction factor. As discussed in \cite{Singh2020_MNRAS_491_51}, the product of all these correction factors has an uncertainty of $\sim \, 6\%$ that is dominated by the uncertainty of $f_{\rm bias}$. Thus, our measured lensing signals have an overall systematic uncertainty of $6\%$. Note that this uncertainty in the normalisation should be independent of scale $r_{\rm p}$ or stellar mass of the sample. Finally, $\Delta\Sigma_{\rm L}$ and $\Delta\Sigma_{\rm R}$ are the raw, uncorrected measurements of the excess surface density for the lenses and a set of random points, respectively. Subtracting the signal around random points can mitigate residual additive systematic biases in the lensing signal and reduce the overall statistical uncertainty \citep{Singh2017_MNRAS_471_3827}. The raw lensing amplitudes are calculated via
\begin{equation}
    \Delta\Sigma_{\rm L} = \frac{\sum_{\rm L} \sum_{\rm S} w_{\rm LS} e_{\rm t} \Delta\Sigma_{\rm crit} (z_{\rm L}, z_{\rm S})}{\sum_{\rm R} \sum_{\rm S} w_{\rm RS}},
\end{equation}
where $\sum_{\rm L}$ denotes a sum over lens galaxies, $\sum_{\rm R}$ a sum over equivalent random targets and $\sum_{\rm S}$ a sum that goes over all sources with $z_{\rm S} > z_{\rm L}$ and within a certain projected distance from the lens or random target. Note that in the denominator, the sum is over random source pairs which amounts to applying the correction for boost factor \citep{Sheldon2004_AJ_127_2544}. Additionally, $e_{\rm t}$ is the tangential ellipticity, $\Delta\Sigma_{\rm crit}$ the critical surface density,
\begin{equation}
    \Delta\Sigma_{\rm crit}(z_{\rm L}, z_{\rm S}) = \frac{1}{(1 + z_{\rm L})^2}\frac{c^2}{4\pi G} \frac{D_{\rm A} (z_{\rm S})}{D_{\rm A}(z_{\rm L}, z_{\rm S}) D_{\rm A}(z_{\rm L})},
\end{equation}

\noindent $D_{\rm A}$ the angular diameter distance and $w_{\rm LS}$ the weight assigned to each lens-source pair. We refer the reader to \cite{Singh2020_MNRAS_491_51} and Leauthaud et al. (in prep.) for a detailed discussion of the galaxy-galaxy lensing computation.

We calculate the galaxy-galaxy lensing signal in $14$ logarithmic bins in $r_{\rm p}$ going from $0.1 \, h^{-1} \mathrm{Mpc}$ to $63 \, h^{-1} \mathrm{Mpc}$, the same bins as for clustering. Similarly, uncertainties are derived from jackknife resampling of $68$ regions. We apply the same Gaussian smoothing to the covariance matrix as for $w_{\rm p}$ to de-noise the covariance estimate. We ignore the cross covariance between clustering and galaxy-galaxy lensing measurements.

\section{Modeling}
\label{sec:modeling}

To make predictions for galaxy clustering and galaxy-galaxy lensing, we directly populate dark matter-only simulations with galaxies. This approach, in contrast to empirical halo models \citep[see e.g.][]{vandenBosch2013_MNRAS_430_725}, is necessary for percent-level accurate predictions in the highly non-linear regime \citep[see e.g.][]{Reid2014_MNRAS_444_476,Leauthaud2017_MNRAS_467_3024,Saito2016_MNRAS_460_1457,McClintock2019_ApJ_872_53, McClintock2019_arXiv_1907_3167}. For this work, we use simulations from the publicly available Abacus simulation suite \citep{Garrison2018_ApJS_236_43}. Specifically, we use the $z=0.3$ outputs from the $20$ \texttt{AbacusCosmos\_720box\_planck} simulation runs. The output redshift is close to the mean redshift of the BOSS LOWZ sample analysed, $z=0.285$. The cosmology used in these simulations is characterised by $H_0 = 67.26 \, \mathrm{km} \, \mathrm{s}^{-1} \, \mathrm{Mpc}^{-1}$, $\Omega_{m, 0} = 0.3142$ and $\sigma_8 = 0.830$. Particularly, $S_8 = 0.849$ is on the high end of the \cite{PlanckCollaboration2020_AA_641_6} analysis where $S_8 = 0.832 \pm 0.013$ (TT,TE,EE+lowE+lensing). As shown in section~\ref{subsec:cosmology}, we expect that using the Planck values would lower the lensing prediction by less than $5\%$, without a strong mass or scale dependence. Thus, it would not qualitatively change the results of this work. Halos in the simulation are identified with the {\sc ROCKSTAR} halo finder \citep{Behroozi2013_ApJ_762_109}. Additionally, we use a random $0.5\%$ subset of all simulation particles to probe the underlying matter density field.

\subsection{Galaxy-Halo Connection}
\label{subsec:galaxy-halo_connection}

There exist several methods to populate dark matter-only simulations with galaxies \citep[see][for a review]{Wechsler2018_ARAA_56_435}. These methods include semi-analytic models, semi-empirical models and subhalo abundance matching models. In this work, we populate dark matter haloes in the simulation according to a Halo Occupation Distribution (HOD) model. Compared to the other three methods, HOD models have the greatest flexibility and allow us to make the smallest amount of intrinsic assumptions about the galaxy-halo connection. More specifically, we use a conditional stellar mass function (CSMF) parameterisation which allows us to predict the abundance of galaxies as well as their clustering and lensing properties as a function of stellar mass. We note that \cite{Leauthaud2017_MNRAS_467_3024} have shown that the exact choice of the galaxy-halo connection model does not have a strong impact on the lensing prediction at fixed clustering.
  
In our CSMF framework, each isolated halo can host two types of galaxies: central galaxies are placed at the centre of haloes and satellite galaxies orbit inside the gravitational potential well. We assume that the average number $dN$ of galaxies with a stellar mass in the range $\log M_\star \pm d \log M_\star / 2$ living in a halo of mass $M_{\rm h}$ is given by
\begin{equation}
    \frac{dN}{d \log M_\star} (M_\star | M_{\rm h}) = \Phi_{\rm c} (M_\star | M_{\rm h}) + \Phi_{\rm s} (M_\star | M_{\rm h}).
\end{equation}
We furthermore assume that the number of centrals follows a Bernoulli distribution, i.e. the number can only be $0$ or $1$, and the number of satellites follows a Poisson distribution.

The central CSMF $\Phi_{\rm c}$ is given by a log-normal distribution
\begin{equation}
    \Phi_{\rm c} (M_\star | M_{\rm h}) = \frac{1}{\sqrt{2 \pi \sigma_{M_\star}^2}} \exp \left[ - \frac{(\log \tilde{M}_\star (M_{\rm h}) / M_\star)^2}{2 \sigma_{M_\star}^2} \right],
\end{equation}
where the characteristic stellar mass is parameterised by the stellar-to-halo mass relation (SHMR),
\begin{equation}
    \tilde{M}_\star (M_{\rm h}) = M_{\star, 0} \frac{(M_{\rm h} / M_{{\rm h}, 1})^{\gamma_1}}{\left[ 1 + (M_{\rm h} / M_{{\rm h}, 1}) \right]^{\gamma_1 - \gamma_2}}.
\end{equation}
Overall, we have five parameters, $\sigma_{M_\star}$, $\log M_{\star, 0}$, $\log M_{{\rm h}, 1}$, $\gamma_1$ and $\gamma_2$, parameterising the central galaxy occupation. However, we fix $\gamma_1 = 4.0$ because it is virtually unconstrained at the high stellar masses we are probing.

Similarly, the satellite CSMF is given by
\begin{equation}
    \begin{split}
    \Phi_{\rm s} (M_\star | M_{\rm h}) = &\phi_{\rm s} (M_{\rm h}) \, (\ln 10) \, \left( \frac{M_\star}{M_\star^\dag (M_{\rm h})} \right)^{\alpha_{\rm s} + 1}\\
    &\exp \left[ -10^{\delta_{\rm s}} \left( \frac{M_\star}{M_\star^\dag (M_{\rm h})} \right)^{2} \right].
    \end{split}
\end{equation}
where
\begin{equation}
    \log \phi_{\rm s} (M_{\rm h}) = b_0 + b_1 \log M_{\rm h} / (10^{12} h^{-1} M_\odot)
\end{equation}
and
\begin{equation}
    \log M_\star^\dag (M_{\rm h}) = \log \tilde{M}_\star (M_{\rm h}) - 0.25
\end{equation}
These definitions follow the parameterisations used in \cite{Yang2008_ApJ_676_248}, \cite{Cacciato2009_MNRAS_394_929} and \cite{Lange2018_MNRAS_473_2830} and have four free parameters: $\alpha_{\rm s}$, $\delta_{\rm s}$, $b_0$ and $b_1$.

The above model describes the abundance and stellar masses of all galaxies. However, only a subset of all galaxies, luminous red galaxies (LRGs), receive spectroscopic redshifts in BOSS. Thus, we need to model this selection, as well. We assume that the probability $c$ for a galaxy to obtain a spectroscopic redshift in in BOSS LOWZ depends on both its stellar mass $M_\star$ and its halo mass $M_{\rm h}$ in the following way:
\begin{equation}
    c (M_\star, M_{\rm h}) = \frac{1}{2} \mathrm{erfc} \left( - \frac{\log M_\star^{\alpha_\Gamma} \tilde{M}_\star^{1 - \alpha_\Gamma} (M_{\rm h}) M_{\star, \Gamma}^{-1}}{\sigma_\Gamma} \right).
\end{equation}
The parameters $M_{\star, \Gamma}$ and $\sigma_\Gamma$ determine the mass and rate at which the completeness changes from $0$ to $1$. Additionally, the parameter $\alpha_\Gamma \in [0, 1]$ determines how much the completeness depends on stellar mass versus halo mass. Specifically, for $\alpha_\Gamma = 1$ it depends purely on stellar mass and for $\alpha_\Gamma = 0$ on halo mass only. This definition generalizes the one used in \cite{Guo2018_ApJ_858_30} by introducing a possible halo mass dependence. Such a dependence is necessary if galaxy properties determining BOSS selection cuts, i.e. luminosity and colour, correlate with halo mass at fixed observed stellar mass \citep{Saito2016_MNRAS_460_1457, Berti2021_AJ_161_49}. In principle, we could vary the three free parameters for centrals and satellites independently. However, we set $\alpha_\Gamma = 1$ for satellites since those parameters would be largely degenerate with the satellite occupation parameters and use the same $\sigma_\Gamma$ for centrals and satellites. Overall, we have four free parameters describing the incomplentess of the BOSS LOWZ sample: $M_{\star, \Gamma, \rm c}$, $M_{\star, \Gamma, \rm s}$, $\sigma_\Gamma$ and $\alpha_\Gamma$

Finally, we assume satellites inside a dark matter halo to be distributed according to an NFW profile \citep{Navarro1997_ApJ_490_493},
\begin{equation}
    n (r) \propto \frac{1}{\frac{\eta r}{r_s} \left( 1 + \frac{\eta r}{r_s} \right)^2},
\end{equation}
where $\log \eta$ is a free parameter that regulates the spatial bias of satellites and $r_s$ is the (dark matter) scale radius of the halo. We vary $\eta$ independently in the three different stellar mass bins for which we measure galaxy clustering and galaxy-galaxy lensing. Given the effect of mass segregation \citep{vandenBosch2016_MNRAS_455_158}, one would expect $\eta$ to increase with stellar mass. Overall, we have $15$ free parameters describing the galaxy-halo connection.

\subsection{Mock Observables}

We use {\sc{halotools}} \citep{Hearin2017_AJ_154_190} to create mock galaxy populations from halo catalogues and the  parameterised galaxy-halo connection. Furthermore, we use the same software package to create mock observables to be compared from these mock galaxy catalogues. For each parameter choice of the galaxy-halo connection, we calculate mock observables, i.e. $w_{\rm p}$ and $\Delta\Sigma$, by averaging the results from all $20$ simulation boxes. We refer the reader to \cite{Lange2019_MNRAS_488_5771} for a detailed discussion of the equations underlying the estimation of $\Delta\Sigma$ from simulations. Finally, we use a pre-computation algorithm\footnote{\url{https://github.com/johannesulf/TabCorr}} \citep{Reid2014_MNRAS_444_476,Zheng2016_MNRAS_458_4015, Lange2019_MNRAS_488_5771} to speed up the calculation of mock observables. The main idea is to compute halo auto- and cross-correlation functions, i.e. $w_{\rm p}$, as well as the halo-matter cross-correlation functions, i.e. $\Delta\Sigma$, as a function of halo mass. To this end, we use $100$ bins in halo mass going from $\log M_h / h^{-1} M_\odot = 12.0$ to $15.4$, the highest halo mass in the simulations. These correlation functions can then be convolved with the galaxy occupation as a function of halo mass to predict the clustering properties and lensing properties of galaxy samples. Thus, to make predictions for galaxies, one does not need to analyse the positions of individual galaxies and matter particles, thereby greatly reducing the computational cost \citep{Zheng2016_MNRAS_458_4015}.

We note that when calculating the expected galaxy-galaxy lensing signal, we place satellites into random positions in the host halo according to an NFW profile. However, this ignores the fact that satellites are hosted by subhalos that are themselves density peaks inside the host dark matter halo. This additional subhalo lensing term has been measured in observations \citep[see e.g.][]{Li2016_MNRAS_458_2573, Sifon2018_MNRAS_478_1244} but its exact contribution for our lensing predictions cannot be predicted a priori because it depends on the relation between observed stellar and subhalo mass. Following the model in \cite{Zu2015_MNRAS_454_1161}, we estimate that the additional contribution of a subhalo lensing term is of the order of $\sim 15\%$ for the low-mass samples and $\sim 5\%$ for the high-mass samples at $r_{\rm p} = 0.1 \, \mathrm{Mpc} / h$. However, the effect should fall off steeply with $r_{\rm p}$ and be negligible at $r_{\rm p} \gtrsim 0.5 \, \mathrm{Mpc} / h$. Ultimately, accounting for the subhalo lensing part would only increase our lensing prediction. Thus, ignoring this effect is a conservative assumption regarding the finding that lensing is low.

\section{Results}
\label{sec:results}

Our analyses follows the general approach employed in \cite{Leauthaud2017_MNRAS_467_3024}. First, we fit a model for the galaxy-halo connection to the SMF and clustering properties of galaxies. This is done separately for the three different measurements corresponding to the three different stellar mass estimates. Afterwards, for each each stellar mass estimate, we study the predictions for the stellar mass-dependent lensing amplitude predictions and compare them against our measurements. Specifically, we want to investigate whether the ratio of observed to predicted lensing amplitude depends on halo mass or radial scale.

\subsection{Galaxy Clustering}

\begin{figure*}
    \centering
    \includegraphics{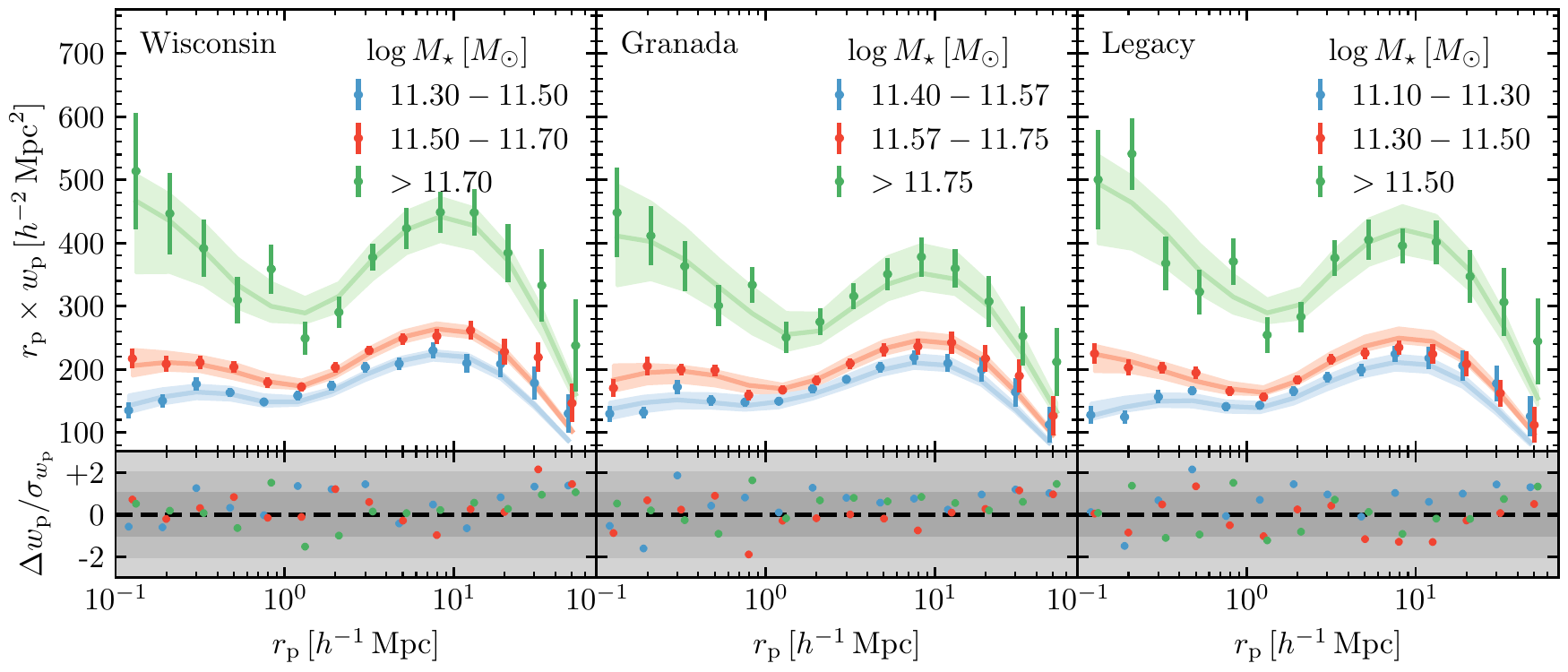}
    \caption{Projected galaxy clustering as a function of comoving projected separation. From left to right, we show the results for the Wisconsin, Granada and Legacy mass estimates, respectively. Different colours correspond to the different stellar mass bins. The upper panel displays the measurements with $1\sigma$ uncertainties as error bars and the bands signify the $95\%$ posterior of the model fitted to the clustering. The lower panels show the difference between the measurements and the best-fit models in terms of $\sigma$. Overall, the Wisconsin and Legacy mass estimates produce a slightly larger correlation of mass with clustering properties than the Granada mass estimates.}
    \label{fig:wp}
\end{figure*}

In the upper panels of Fig.~\ref{fig:wp}, we show the projected galaxy clustering measurements $w_{\rm p, obs}$ for the different stellar mass selected samples. For all three stellar mass estimates, we find that higher stellar masses result in larger clustering amplitudes on all scales. This is expected because of the correlations between stellar mass and halo mass as well as halo mass and clustering. When comparing the three different stellar mass estimates, we find that the Wisconsin mass estimates lead to the strongest clustering differences between different stellar mass samples, followed by the Legacy mass estimates and finally the Granada mass estimates.

\begin{table*}
    \centering
    \begin{tabular}{c|cc|ccc}
        \multirow{2}{*}{Parameter} & \multicolumn{2}{c}{Prior} & \multicolumn{3}{c}{Posterior} \\
        & Minimum & Maximum & Wisconsin & Granada & Legacy\\
        \hline
        $\log M_{\star, 0}$ & 10.0 & 12.0 & $11.06_{-0.28}^{+0.31}$ & $11.37_{-0.39}^{+0.12}$ & $10.73_{-0.26}^{+0.29}$ \\
        $\log M_{{\rm h}, 1}$ & 11.0 & 14.0 & $11.84_{-0.49}^{+0.39}$ & $12.12_{-0.54}^{+0.13}$ & $11.72_{-0.45}^{+0.36}$ \\
        $\gamma_2$ & 0.05 & 0.4 & $0.293_{-0.098}^{+0.055}$ & $0.173_{-0.055}^{+0.124}$ & $0.329_{-0.081}^{+0.043}$ \\
        $\sigma_{\log M_\star}$ & 0.1 & 0.25 & $0.132_{-0.011}^{+0.011}$ & $0.171_{-0.012}^{+0.012}$ & $0.149_{-0.014}^{+0.014}$ \\
        $b_0$ & -3.5 & 1.5 & $0.56_{-0.86}^{+0.46}$ & $-2.54_{-0.64}^{+2.74}$ & $0.63_{-0.68}^{+0.44}$ \\
        $b_1$ & -2.0 & 3.0 & $-0.04_{-0.23}^{+0.37}$ & $1.25_{-1.10}^{+0.22}$ & $-0.10_{-0.20}^{+0.31}$ \\
        $2 + \alpha_{\rm s}$ & -3.0 & 2.0 & $-2.30_{-0.49}^{+1.04}$ & $1.16_{-3.19}^{+0.70}$ & $-2.38_{-0.37}^{+0.73}$ \\
        $\sigma_\Gamma$ & 0.01 & 0.3 & $0.044_{-0.023}^{+0.026}$ & $0.049_{-0.025}^{+0.030}$ & $0.085_{-0.020}^{+0.017}$ \\
        $\log M_{\Gamma, \rm c}$ & 10.0 & 12.0 & $11.380_{-0.011}^{+0.012}$ & $11.439_{-0.015}^{+0.010}$ & $11.188_{-0.013}^{+0.013}$ \\
        $\log M_{\Gamma, \rm s}$ & 10.0 & 12.0 & $11.428_{-0.013}^{+0.010}$ & $11.424_{-0.978}^{+0.071}$ & $11.244_{-0.016}^{+0.016}$ \\
        $\alpha_\Gamma$ & 0.0 & 1.0 & $0.33_{-0.15}^{+0.18}$ & $0.78_{-0.30}^{+0.16}$ & $0.66_{-0.15}^{+0.17}$ \\
        $\delta_{\rm s}$ & -1.0 & 1.0 & $-0.53_{-0.24}^{+0.23}$ & $-0.41_{-0.31}^{+0.12}$ & $-0.72_{-0.18}^{+0.23}$ \\
        $\eta_1$ & 0.5 & 2.0 & $0.611_{-0.080}^{+0.187}$ & $0.585_{-0.066}^{+0.130}$ & $0.579_{-0.056}^{+0.136}$ \\
        $\eta_2$ & 0.5 & 2.0 & $1.05_{-0.25}^{+0.30}$ & $0.84_{-0.23}^{+0.40}$ & $1.60_{-0.30}^{+0.26}$ \\
        $\eta_3$ & 0.5 & 2.0 & $1.74_{-0.31}^{+0.17}$ & $1.57_{-0.39}^{+0.30}$ & $1.76_{-0.25}^{+0.16}$ \\
    \end{tabular}
    \caption{Prior and posterior constraints on galaxy-halo connection parameters for the three different stellar mass estimates. All priors are chosen to be flat.}
    \label{tab:posterior}
\end{table*}

We now fit the galaxy-halo connection model described in section~\ref{subsec:galaxy-halo_connection} to the observed SMF $\Phi_{\rm obs}$ and projected clustering $w_{\rm p, obs}$. This is done for each of the three stellar mass estimates separately. We assume flat priors for all parameters as listed in Table~\ref{tab:posterior} and a multi-variate Gaussian likelihood, i.e.
\begin{equation}
    \ln \mathcal{L} = - \frac{\chi_{\rm SMF}^2 + \chi_{w_{\rm p}}^2}{2},
\end{equation}
where
\begin{equation}
    \chi_{\rm SMF}^2 = \sum_i \frac{\left( \Phi_{\rm obs}(M_{\star, i}) - \Phi_{\rm mod}(M_{\star, i}) \right)^2}{\left( 0.05 \, \Phi_{\rm obs}(M_{\star, i}) \right)^2} \, ,
\end{equation}
reflecting a $5\%$ uncorrelated error on the SMF, as discussed earlier, and
\begin{equation}
    \chi_{w_{\rm p}}^2 = (w_{\rm p, obs} - w_{\rm p, mod})^T C_{w_{\rm p}}^{-1} (w_{\rm p, obs} - w_{\rm p, mod}).
\end{equation}

We use the nested sampling \citep{Skilling2004_AIPC_735_395} code {\sc MultiNest} \citep{Feroz2008_MNRAS_384_449, Feroz2009_MNRAS_398_1601, Feroz2019_OJAp_2_10} to evaluate the posterior of galaxy-halo connection parameters. We use $5000$ live points, a target efficiency of $5\%$ and a stopping criterion of $\Delta\ln \mathcal{Z} = 10^{-3}$. Constant efficiency mode is turned off.

\begin{figure*}
    \centering
    \includegraphics{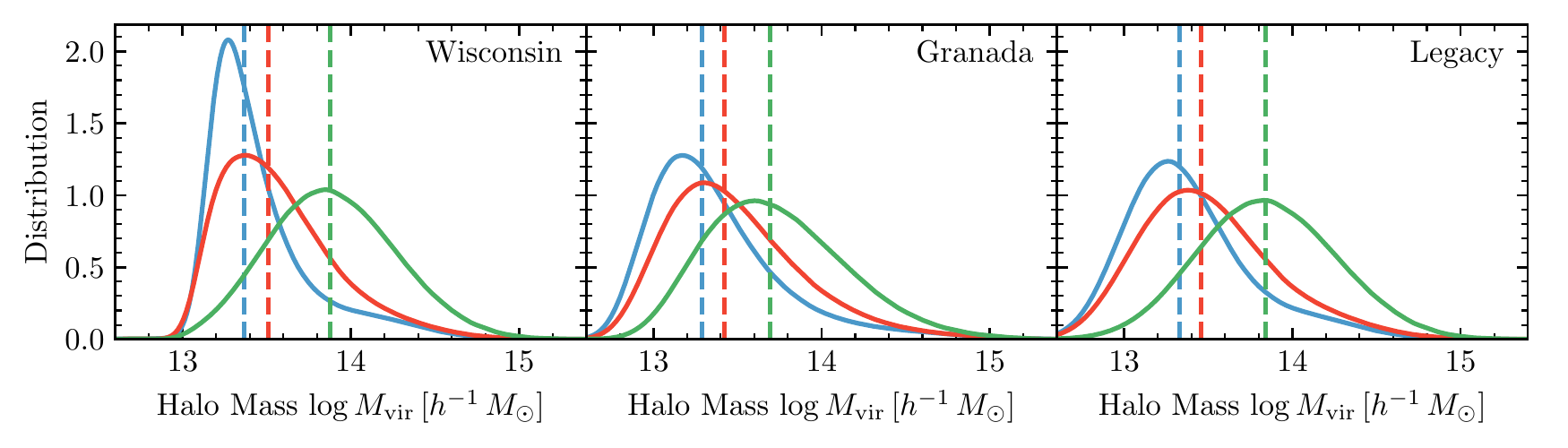}
    \caption{Clustering-based predictions on the halo mass distribution for the different stellar mass bins. The distribution is predicted from the best-fitting model with respect to the SMF and galaxy clustering data. Colours have the same meaning as in Fig. \ref{fig:wp}. Vertical dashed lines denote the median host halo mass of each galaxy sample. As expected, the models fitted to the Wisconsin and Legacy stellar masses predict a slightly larger difference in host halo masses between different stellar mass subsamples.}
    \label{fig:halo_mass_dist}
\end{figure*}

The clustering prediction of the galaxy-halo connection model is shown in the upper panel of Fig.~\ref{fig:wp} with bands denoting $95\%$ uncertainty ranges. Similarly, the lower panels display the difference between the best-fit model and the observations in units of the observational uncertainty. Overall, the model is able to qualitatively predict the clustering amplitudes for all three stellar mass estimates. The $\chi^2$ value is $38$, $33$ and $39$ for $47$ data points and $15$ free parameters for the Wisconsin, Granada and Legacy masses, respectively. In our galaxy-halo connection model, the different clustering properties of the three stellar mass estimates are largely explained by different scatter of stellar mass at fixed halo mass. The scatter is $\sigma_{M_\star} = 0.132_{-0.011}^{+0.011}$, $0.171_{-0.012}^{+0.012}$ and $0.149_{-0.014}^{+0.014}$ for the Wisconsin, Granada and Legacy stellar masses, respectively. This follows the trend observed in the clustering with the Wisconsin (Granada) masses having the strongest (weakest) correlation of stellar mass with clustering and smallest (largest) $\sigma_{M_\star}$. For all three stellar mass estimates, we also find that the model favours mass segregation, i.e. $\eta_1 < \eta_2 < \eta_3$. In Fig.~\ref{fig:halo_mass_dist}, we show the predicted host halo mass distributions for each of the three best-fit models. As expected, the Wisconsin and Legacy models predict a stronger difference in the halo mass distributions of the three stellar mass bins than the Granada model.

\subsection{Galaxy-Galaxy Lensing}
\label{subsec:lensing}

\begin{figure*}
    \centering
    \begin{subfigure}
        \centering
        \includegraphics{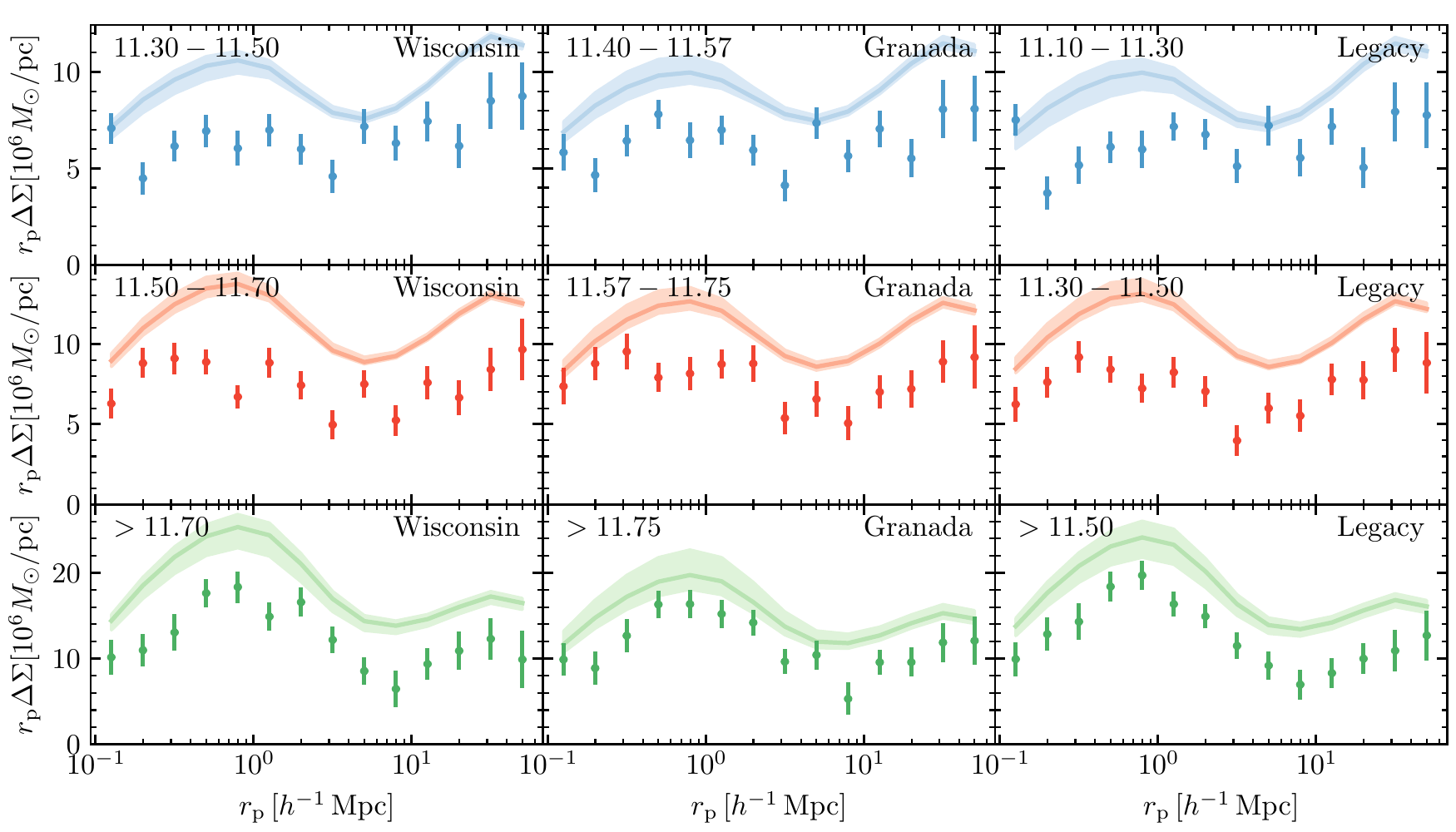}
    \end{subfigure}
    \begin{subfigure}
        \centering
        \includegraphics{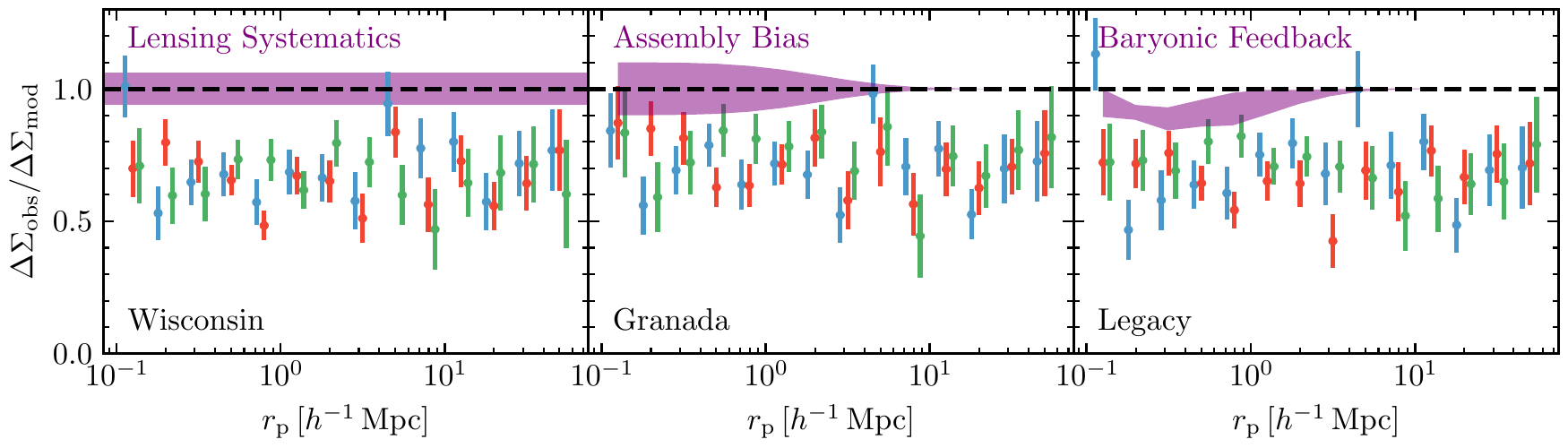}
    \end{subfigure}
    \caption{The galaxy-galaxy lensing signal of different stellar mass subsamples and the ratio of observed to predicted lensing amplitude. In the upper panel we compare predictions (bands, $95\%$ uncertainty) from clustering and observations (error bars, $68\%$ uncertainty). The top left label indicates the stellar mass range and the top right label the stellar mass estimate. In the bottom panel, we show the ratio of observed to predicted lensing signal. The $68\%$ error bars include both observational uncertainties and model uncertainties. In the bottom panel, the results for different stellar mass subsamples are offset in the $x$-axis for clarity. Generally, the lensing signal is over-predicted by $\sim 35\%$ on all scales, for all stellar masses, and all stellar mass estimates. Finally, in the lower panel we also show the potential contribution from lensing systematics, galaxy assembly bias and baryonic feedback.}
    \label{fig:ds}
\end{figure*}

As discussed in \cite{Leauthaud2017_MNRAS_467_3024}, fitting galaxy-halo models to small-scale clustering data provides precise predictions for the lensing amplitude if cosmology is kept fixed and assembly bias and baryonic feedback are ignored. In the upper panels of Fig.~\ref{fig:ds}, we show as bands the $95\%$ uncertainty predictions for the galaxy-galaxy lensing amplitude in different stellar mass bins. As expected, the galaxy-halo connection models based on clustering predict a positive correlation between the lensing amplitude and the stellar mass of the sample. Similarly, in accordance with the results in the previous subsection, the models for the Wisconsin and Legacy mass estimates predict the widest spread in lensing amplitudes between the different subsamples.

In the same panels, we show as error bars the measurements from cross-correlating our samples with SDSS galaxy shapes. We see that the measurements reproduce the positive correlation of stellar mass and lensing amplitude. However, for all three stellar mass estimates, the lensing amplitude is significantly over-predicted for almost all stellar mass bins and on all scales. In the bottom panels of the same figure we show the ratio of observed to predicted lensing signal. The uncertainties include both the observational uncertainties as well as model uncertainties from fitting the galaxy-halo model to the clustering data. We see highly significant deviations from the expected unity ratio and find $\Delta\Sigma_{\rm obs} / \Delta\Sigma_{\rm mod} \sim 0.65 - 0.70$ instead. We also show, as a guidance, the $6\%$ systematic uncertainty coming from the lensing systematics, the $1\sigma$ uncertainty from not modelling galaxy assembly bias \citep{Lange2019_MNRAS_488_5771} and the impact of baryonic feedback whereby the band show the range between the predictions from Illustris and IllustrisTNG \citep{Lange2019_MNRAS_490_1870}.

First, we look for a scale-dependence of the ratio $f = \Delta\Sigma_{\rm obs} / \Delta\Sigma_{\rm mod}$. To this end, we fit the data shown in the lower panel of Fig~\ref{fig:ds} with a simple linear model, $f = a + b \log r_{\rm p}$. Irrespective of stellar mass estimate or bin, we find $b$ to be constrained to within $\sim \pm 0.04$. However, for all the nine samples we analysed, $b$ is consistent with $0$, i.e. no scale dependence, to within $\lesssim 1.5 \sigma$. Thus, we do not find any evidence for a strong scale dependence of the ratio of observed to predicted lensing signal. In the following, we will average $f$ over all scales to study the mass dependence of the signal. The scale-averaged lensing ratios are shown in Fig.~\ref{fig:cosmology} for all three different stellar mass estimates and samples. Similar to the scale dependence, we do not find evidence for a mass dependence for any of the three stellar mass estimates; the ratio is always consistent with $f \sim 0.65$.

\section{Discussion}
\label{sec:discussion}

In the previous section, we studied the scale and stellar mass dependence of the galaxy clustering and galaxy-galaxy lensing amplitude. Here, we will discuss implications for the lensing-is-low tension and cosmology as well as scatter in the SHMR and different stellar mass estimates.

\subsection{Lensing is Low}
\label{subsec:lensing_is_low}

\begin{figure}
    \centering
    \includegraphics{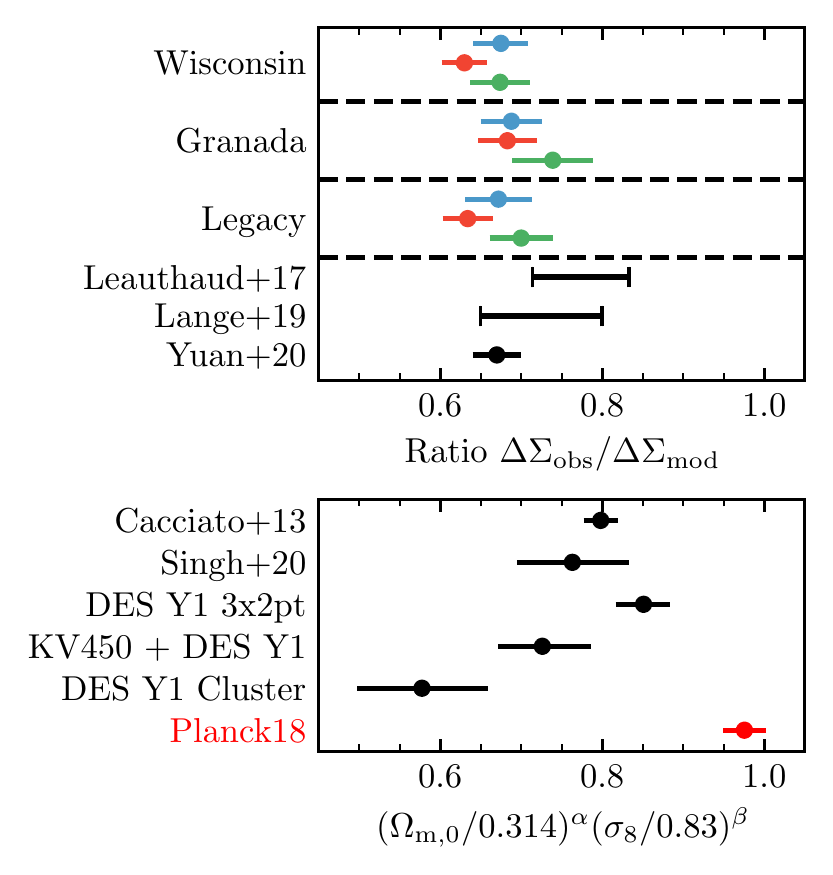}
    \caption{The scale-averaged ratio of observed to predicted lensing signal (top panel) and constraints on cosmological parameters from the literature (bottom panel). In the upper panel, we compare our results on the ratio of predicted to observed lensing signal for different stellar mass bins against those of \protect\cite{Leauthaud2017_MNRAS_467_3024}, \protect\cite{Lange2019_MNRAS_488_5771} and \protect\cite{Yuan2020_MNRAS_493_5551}. In the lower panel we show constraints on $\Omega_{{\rm m}, 0}^\alpha \sigma_8^\beta$ \protect\citep{Cacciato2013_MNRAS_430_767, Singh2020_MNRAS_491_51, Abbott2018_PhRvD_98_3526, Abbott2020_PhRvD_102_3509, Asgari2020_AA_634_127, PlanckCollaboration2020_AA_641_6} divided by the values in the Abacus Planck simulations. For $\alpha = 1.0$ and $\beta = 1.25$ and in the absence of galaxy assembly bias and baryonic feedback, this ratio should be roughly comparable to the ratio of observed to predicted lensing signal in our analysis. Error bars denote $68\%$ uncertainties and for \protect\cite{Leauthaud2017_MNRAS_467_3024} and \protect\cite{Lange2019_MNRAS_488_5771} we show rough ranges. See the text for details.
    }
    \label{fig:cosmology}
\end{figure}

\begin{figure}
    \centering
    \includegraphics{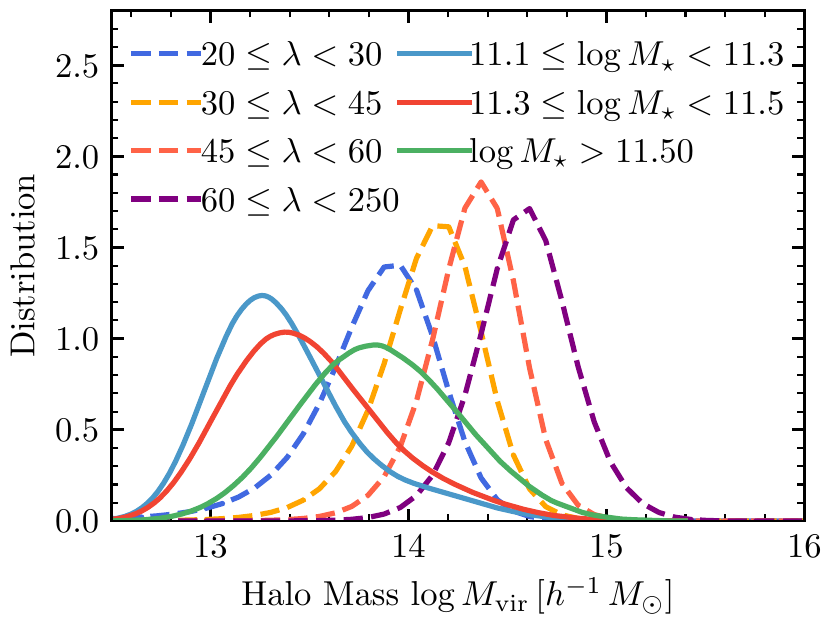}
    \caption{The inferred host halo mass distribution of different galaxy and cluster samples. The solid lines show the best-fit inferred halo masses for the BOSS LOWZ galaxies in this work, based on modelling their clustering properties and the Legacy stellar mass estimates. Dashed lines show the inferred halo mass distribution of clusters in different richness bins from the DES cluster cosmology analysis \protect\citep{Abbott2020_PhRvD_102_3509}. These estimates are based on modelling cluster counts and weak lensing signals. Our most massive galaxy bin corresponds to the lowest DES richness bin.}
    \label{fig:halo_mass_dist_des}
\end{figure}

Our results in section~\ref{subsec:lensing} show ubiquitous findings of a lensing-is-low like tension with respect to cosmological parameters from the \cite{PlanckCollaboration2020_AA_641_6} CMB analysis. This confirms earlier results by \cite{Leauthaud2017_MNRAS_467_3024}, \cite{Lange2019_MNRAS_488_5771} and \cite{Yuan2020_MNRAS_493_5551} finding a lensing underprediction in BOSS when fixing clustering but increases the signal-to-noise ratio compared to these studies through the use of galaxy-galaxy lensing from SDSS over CS82 and CFHTLenS. In the upper panel of Fig.~\ref{fig:cosmology}, we compare our finding for $\Delta\Sigma_{\rm obs} / \Delta\Sigma_{\rm mod}$ against these other works from the literature.

Our results can be directly compared to \cite{Yuan2020_MNRAS_493_5551} where the authors fit the clustering and lensing properties of BOSS CMASS galaxies. They assume the same cosmological parameters for the modelling and infer $\Delta\Sigma_{\rm obs} / \Delta\Sigma_{\rm mod} = 0.67 \pm 0.03$, in good agreement with our results. Similarly, they find that this ratio is consistent for all scales analysed, albeit with large uncertainties, especially for $r_{\rm p} \gtrsim 5 h^{-1} \mathrm{Mpc}$. \cite{Leauthaud2017_MNRAS_467_3024} use the same lensing data as \cite{Yuan2020_MNRAS_493_5551} but compare models fitted to galaxy clustering from different studies \citep{Reid2014_MNRAS_444_476, Saito2016_MNRAS_460_1457, RodriguezTorres2016_MNRAS_460_1173, Alam2017_MNRAS_470_2617}. Unfortunately, the different studies do not all assume the same cosmology, making a direct comparison to our results difficult. Overall, the authors find $\Delta\Sigma_{\rm obs} / \Delta\Sigma_{\rm mod} \sim (1.2 - 1.4)^{-1}$ and no strong evidence for a scale dependence, in qualitative agreement with our results. Finally, \cite{Lange2019_MNRAS_488_5771} analyse $\Delta\Sigma_{\rm obs} / \Delta\Sigma_{\rm mod}$ for the majority of galaxies in BOSS LOWZ and CMASS using lensing data from CFHTLenS. When analysing the entire sample, they find a $\sim 3 \sigma$ detection that $\Delta\Sigma_{\rm obs} / \Delta\Sigma_{\rm mod}$ increases at larger radii with a strength compatible with our results in section~\ref{subsec:lensing}. However, the results in \cite{Lange2019_MNRAS_488_5771} are driven by scales $\lesssim 5 \, h^{-1} \, \mathrm{Mpc}$ as larger scales have much larger uncertainty. On small scales, $r_{\rm p} \lesssim 3 \, h^{-1} \mathrm{Mpc}$, where the signal-to-noise ratio is the largest, they find $\Delta\Sigma_{\rm obs} / \Delta\Sigma_{\rm mod} = 0.65 - 0.80$. Note that \cite{Lange2019_MNRAS_488_5771} utilise an analytic halo model when making clustering and lensing predictions. This could lead to inaccuracies in the predictions for $\Delta\Sigma_{\rm obs} / \Delta\Sigma_{\rm mod}$ whereas here we use direct mock population and so our current predictions are more accurate across all radial scales.

Recently, \cite{Zu2020_arXiv_2010_1143} claimed that the lensing-is-low tension can be solved on small scales for both the CMASS sample analysed in \cite{Leauthaud2017_MNRAS_467_3024} and \cite{Yuan2020_MNRAS_493_5551} and a LOWZ sub-sample very similar to the one studied in this work. The author can fit the large-scale clustering and lensing on small scales by predicting a large fraction of satellites in both galaxies samples. Theoretically, this works in reducing the small-scale lensing signal because satellite are off-centred from dark matter halo core and thereby have a smaller small-scale lensing amplitude than centrals at the same halo mass and large-scale bias. However, the fraction of satellites in many studies is tightly constrained by observations that were not studied in \cite{Zu2020_arXiv_2010_1143}, like the projected galaxy clustering down to $0.1 \, h^{-1} \mathrm{Mpc}$ or anisotropic clustering \citep[see e.g.][]{Reid2014_MNRAS_444_476, Guo2015_MNRAS_446_578, Saito2016_MNRAS_460_1457}, or direct counts in clusters (e.g., Bradshaw et al in prep). For example, the best-fit model of \cite{Zu2020_arXiv_2010_1143} implies a satellite fraction of $f_{\rm sat} \sim 0.5 - 0.8$ for CMASS, significantly higher than the constraints from \cite{Reid2014_MNRAS_444_476}, \cite{Guo2015_MNRAS_446_578}, and \citet[][]{Saito2016_MNRAS_460_1457} placing it at $f_{\rm sat} \sim 0.1 \pm 0.03$. 

Our new results provide meaningful constraints on the scale dependence of the lensing-is-low tension. Overall, we find no evidence for a strong scale dependence of the ratio of observed to predicted lensing signal. As discussed in \cite{Lange2019_MNRAS_488_5771}, changes to the cosmological parameters, particularly $S_8$, tend to change the lensing predictions on all scales without a very strong scale dependence. On the other hand, galaxy assembly bias and baryonic feedback have a stronger scale dependence where the impact is limited to $r_{\rm p}  \lesssim 5 \, h^{-1} \mathrm{Mpc}$ and $\lesssim 1 \, h^{-1} \mathrm{Mpc}$, respectively. We note that both effects are not modelled in our analysis as our predictions are based on collisionless dark matter-only simulations and our model for the galaxy-halo connection postulates that galaxy occupation depends on halo mass only. As discussed by \cite{Lange2019_MNRAS_488_5771} and \cite{Yuan2020_MNRAS_493_5551}, both effects can likely alleviate but not completely explain the lensing-is-low tension on small scales. Similarly, they are unable to explain the lensing tension on larger scales. Thus, it remains difficult to resolve the lensing-is-low result without a change in cosmological parameters. On the other hand, the absence of a strong scale dependence to the lensing-is-low tension likely also places interesting constraints on models of baryonic feedback. While baryonic feedback typically impacts the lensing signal at the level of $\sim 10\%$ \citep{Leauthaud2017_MNRAS_467_3024, Lange2019_MNRAS_488_5771}, its strength can vary widely between different feedback implementations \citep{vanDaalen2020_MNRAS_491_2424}. Thus, our lensing data might be able to rule out very energetic baryonic feedback models. Particularly, it will be interesting to compare such constraints to more direct constraints from probing gas physics via the Sunyaev-Zeldovich effect \citep{Amodeo2020_arXiv_2009_5558}. We leave such an analysis to future work.

We also find no evidence for a strong stellar mass dependence of the lensing-is-low signal. Through the correlation of stellar mass and halo mass, this also implies the absence of a strong halo mass dependence in the halo mass range $10^{13.3} - 10^{13.9} h^{-1} M_\odot$. The absence of a strong halo mass dependence in the lensing-is-low effect is consistent with both the baryonic feedback models in Illustris and IllustrisTNG or changes in cosmological parameters \citep{Lange2019_MNRAS_488_5771}. However, our results are qualitatively different than those presented in the DES cluster analysis \citep{Abbott2020_PhRvD_102_3509}. Particularly, \cite{Abbott2020_PhRvD_102_3509} find that only clusters with low richness have a lensing amplitude that is strongly over-predicted when one assumes the cosmology from the DES 3x2pt analysis\footnote{The DES 3x2pt analysis \citep{Abbott2018_PhRvD_98_3526} favours $S_8 = 0.773_{-0.020}^{+0.026}$, lower than the best-fit value of the Planck CMB analysis, $S_8 = 0.825 \pm 0.011$ \citep{PlanckCollaboration2020_AA_641_6}. If the Planck CMB cosmological model was assumed, a lensing-is-low like tension would have likely been found in all richness bins, albeit with the strongest finding still in the lowest richness bin.}. In other words, if not explained by systematic errors, the findings by \cite{Abbott2020_PhRvD_102_3509} indicate a strong halo mass dependence. However, we note that our analysis covers a lower halo mass range than the DES cluster cosmology study, as shown in Fig.~\ref{fig:halo_mass_dist_des}. Particularly, only our highest stellar mass bin roughly overlaps with the lowest richness bin in the DES cluster cosmology analysis. In this lowest richness bin, assuming $\Delta\Sigma \propto M_{\rm h}^{2/3}$, the authors find $\Delta\Sigma_{\rm obs} / \Delta\Sigma_{\rm mod} \approx 0.7 - 0.8$. Taking into account that the predicted $\Delta\Sigma_{\rm mod}$ is based on a cosmology with $\sim 10\%$ lower $S_8$, the ratio would likely be lower for the Planck cosmology and likely in the range of our results. Overall, we find that while it is possible that the findings of \cite{Abbott2020_PhRvD_102_3509} are caused by observational systematics, our analysis does not seem inconsistent with their results in the overlapping halo mass range.

\subsection{Cosmology}
\label{subsec:cosmology}

\begin{figure}
    \centering
    \includegraphics{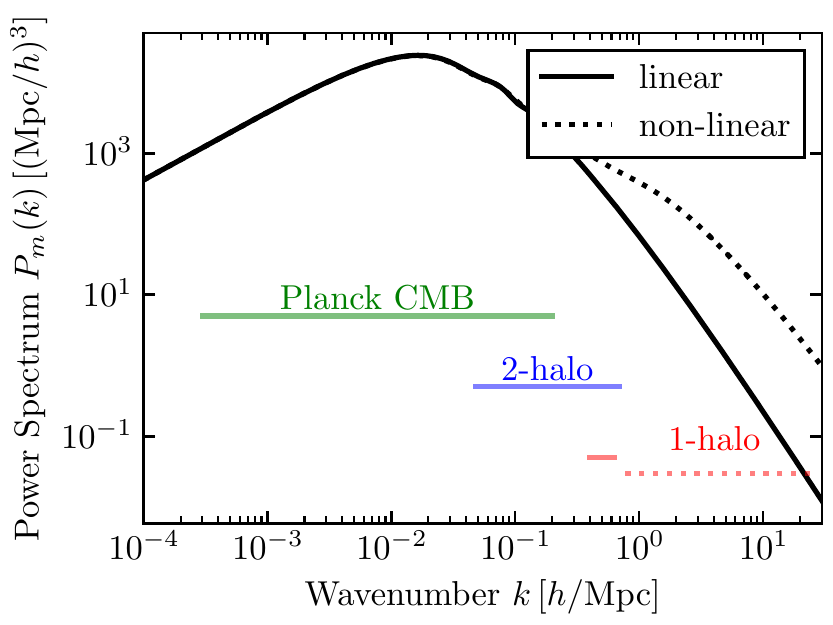}
    \caption{The linear (solid) and non-linear (dotted) matter power spectrum as predicted by the best-fit cosmological parameters of the \protect\cite{PlanckCollaboration2020_AA_641_6} CMB analysis. We also outline the approximate scales probed by different experiments. Following \protect\cite{Chabanier2019_MNRAS_489_2247}, we convert distances $r$ into wavenumbers $k$ by calculating the median of the window function that determines matter fluctuations on scales $r$. For the two-halo regime we use $r = 3.5 - 50 \, h^{-1} \, \mathrm{Mpc}$. For the 1-halo regime we instead show two choices. First, as solid lines, we use the lagrangian radius $r$ of halos of masses in the range $10^{13.3} - 10^{13.9} \, h^{-1} \, M_\odot$. This roughly corresponds to the $k$-range contributing to these haloes in the linear regime. For the dotted lines we use $r = 0.1 - 3.5 \, h^{-1} \, \mathrm{Mpc}$, corresponding to the $k$-range of the 1-halo term in the non-linear regime. Note that both the clustering and lensing probe roughly the same scales.}
    \label{fig:power_spectrum}
\end{figure}

The lensing-is-low tension can be interpreted as evidence for cosmological parameters different than the ones preferred by the Planck CMB analysis. However, running a full cosmological analysis is beyond the scope of this work as it would require us to carefully model baryonic feedback and galaxy assembly bias. Instead, we seek to quantify by how much different cosmologies proposed in the literature can alleviate the tension reported here. We can perform a very rough comparison with other works by noting that the predicted lensing signal at fixed galaxy clustering scales roughly with $\Omega_{{\rm m}, 0}^\alpha \sigma_8^\beta$. In the purely linear regime, the predicted $\Delta\Sigma$ scales with $\Omega_{{\rm m}, 0} \sigma_8$, i.e. $\alpha = \beta = 1$. In the non-linear regime where our signal-to-noise ratio is the highest the relation is more complicated and can vary with scale in the range from $\sim 0.7$ to $\sim 1.1$ for $\alpha$ and $\sim 1.0$ to $\sim 1.5$ for $\beta$ \citep{Yoo2006_ApJ_652_26}. In the following, we will use $\alpha = 1.0$ and $\beta = 1.25$ which is a good approximation for $r_{\rm p} = 1 h^{-1} \mathrm{Mpc}$.

The lower panel of Fig.~\ref{fig:cosmology} demonstrates by roughly how much different proposed cosmologies would lower the predicted lensing signal compared to the prediction from Abacus Planck. \cite{Cacciato2013_MNRAS_430_767} study the luminosity-dependent clustering and lensing properties of galaxies in the main galaxy sample of SDSS to constrain cosmological parameters. Given that \cite{Cacciato2013_MNRAS_430_767} analyse galaxy-galaxy lensing down to the highly non-linear regime and they can explain all the data without the need for baryonic feedback or assembly bias, their results seem at odds with our findings since their inferred cosmology would only lower the lensing prediction by around $20\% \pm 2\%$, not $30\% - 35\%$. Part of the reason could be that \cite{Cacciato2013_MNRAS_430_767} use an analytic halo model to predict galaxy clustering and galaxy-galaxy lensing. We find that this analytic halo model when applied to our data and assumed cosmology tends to underpredict the lensing amplitude at fixed clustering by $\sim 10\%$ compared to the simulation, similar to what was found in \cite{Lange2019_MNRAS_488_5771}.

\cite{Singh2020_MNRAS_491_51} fit the clustering and lensing amplitude of BOSS LOWZ galaxies in the redshift range $0.16 < z < 0.36$ down to scales $r_{\rm p} = 2 \, h^{-1} \mathrm{Mpc}$. Given that they use a very similar sample to ours, it is not surprising that their cosmological constraints would result in a $24\% \pm 7\%$ lower lensing prediction, in rough agreement with what we find. The remaining difference could be caused by their correction for the effects of baryonic feedback that tends to lower the lensing prediction even further. Qualitatively and quantitatively similar statements can be made regarding a comparison with the results of \cite{Wibking2019_MNRAS_484_989} which are based on the same data as \cite{Singh2020_MNRAS_491_51} but use a different analysis method and go down to $r_{\rm p} = 0.6 \, h^{-1} \mathrm{Mpc}$. The DES Y1 3x2pt analysis \citep{Abbott2018_PhRvD_98_3526} concentrates on clustering and lensing on larger scales where the impact of baryonic feedback and details of the galaxy-halo connection is not strong. Their constraints on cosmological parameters imply a $15\% \pm 3\%$ lower lensing prediction. Similarly, the cosmic shear analysis of \cite{Asgari2020_AA_634_127} would result in a roughly $27\% \pm 7\%$ reduction. Finally, the cosmological constraints from the DES Y1 cluster cosmology analysis imply a $42\% \pm 8\%$ reduced lensing prediction.

As discussed above, and shown in Fig.~\ref{fig:cosmology}, both clustering and lensing signals can be fit if we allow for a reduction in $\Omega_{m,0}$ and/or $\sigma_8$, but the required values are in tension with other cosmological measurements. A reduction in either parameter will be in tension with measurements of the CMB within the $\Lambda$CDM cosmology. If we consider a model which results in the reduction in $\Omega_{m,0}$ between the CMB and today (such as a decaying dark matter model) we are confronted with late-time measurements such as the luminosity distance determined by type Ia supernovae which give $\Omega_{m,0} = 0.298 \pm 0.022$ \citep{Scolnic2018_ApJ_859_101}.

Along similar lines, it is of interest to consider modifications of the standard cosmological model which would allow for a non-standard scale dependence. The range of scales probed by the data considered here is shown in Fig.~\ref{fig:power_spectrum}. We convert the distances $r$ into wavenumbers $k$ by calculating the median of the window function that determines matter fluctuations on scales $r$ \citep{Chabanier2019_MNRAS_489_2247}. Using this convention, our observations are sensitive to $\sim 0.05 \, \mathrm{Mpc} / h < k < 20 \, \mathrm{Mpc} / h$. One might ask whether a break to the power spectrum at large $k$ could reconcile the clustering and lensing measurements while still be in agreement with CMB constraints. We note that $\Delta\Sigma$ at fixed clustering roughly scales with $\sigma_8$ and the power spectrum with $\sigma_8^2$. Thus, the change in the power spectrum compared to the $\Lambda$CDM prediction would have to be large, $\gtrsim 30\%$ to lower the lensing prediction significantly. However, the clustering measurements presented here require that this reduction be fairly scale-independent between $0.05 \, h \, {\rm Mpc}^{-1} \lesssim k \lesssim 1 \, h \, {\rm Mpc}^{-1}$. Using the model of \cite{vandenBosch2013_MNRAS_430_725} we checked that a suppression of the power spectrum going from $0\%$ to $30\%$ in the range $k = 0.02 h \, {\rm Mpc}^{-1}$ to $k = 0.1 h \, {\rm Mpc}^{-1}$ provides an insufficient fit to the shape of the projected clustering measurements while not solving the lensing tension on all scales. Thus, a break in the power spectrum would have to occur at even smaller $k$. However, as shown in Fig.~\ref{fig:power_spectrum}, the lower end of these scales overlap with those probed by the CMB ($k\simeq 0.05 h/{\rm Mpc}$ roughly corresponds to a multipole $\ell \simeq 400$), which implies that any pre-recombination modification to the matter power spectrum (see, e.g., ~\citealt{BuenAbad2018_JCAP_01_008}) must be accompanied by a modification to the photon transfer function to compensate. Post-recombination suppression of the matter power spectrum may be achieved by decaying dark matter scenarios, such as the one discussed in \cite{Abellan2020_arXiv_2008_9615}. 

\subsection{Scatter in the Stellar-to-Halo Mass Relation}

The stellar mass scatter in the SHMR is an important probe of galaxy evolution \citep[see e.g.][]{Gu2016_ApJ_833_2, Wechsler2018_ARAA_56_435} as it relates to the stochasticity of star formation and the time-scale of feedback processes \citep[see e.g.][]{Hahn2019_arXiv_1910_1644}. In our analysis we find a scatter of $0.13$ to $0.17 \, \mathrm{dex}$, depending on the stellar mass estimate. This finding is in agreement with other studies which place the scatter at $\sim 0.1 - 0.2 \, \mathrm{dex}$ \citep{Yang2009_ApJ_695_900, More2011_MNRAS_410_210, Leauthaud2012_ApJ_744_159, Zu2015_MNRAS_454_1161, Saito2016_MNRAS_460_1457, Tinker2017_ApJ_839_121, Behroozi2019_MNRAS_488_3143} with the scatter likely being higher for smaller halo masses \citep{Zu2015_MNRAS_454_1161, Lange2019_MNRAS_487_3112, Cao2020_MNRAS_498_5080}.

Our results can be most directly compared to the findings of \cite{Tinker2017_ApJ_839_121}. The authors study the mass-dependent clustering amplitude of BOSS CMASS galaxies, similar to what we perform for LOWZ. However, \cite{Tinker2017_ApJ_839_121} infer a scatter of $0.18_{-0.02}^{+0.01} \, \mathrm{dex}$ for the Wisconsin masses whereas we find $0.13 \pm 0.01 \, \mathrm{dex}$. The difference could be partially explained by the different galaxy samples, i.e. CMASS versus LOWZ, and the fact that \cite{Tinker2017_ApJ_839_121} assume a cosmology with $\Omega_{\rm m} = 0.27$ and $\sigma_8 = 0.82$. On the other hand, \cite{Saito2016_MNRAS_460_1457} studied the anisotropic clustering of BOSS CMASS galaxies using the $M_\star$ estimates of \cite{Bundy2015_ApJS_221_15}, i.e. comparable to our Legacy mass estimates. They infer a $0.10 - 0.14 \, \mathrm{dex}$ scatter between stellar mass and halo $V_{\rm peak}$, the peak maximum circular velocity $V_{\rm max}$ achieved over the life-time of a halo. Given the close correlation between $V_{\rm peak}$ and $M_{\rm vir}$ for field halos, our results of $\sigma_{\log M_\star} = 0.149_{-0.014}^{+0.014}$ are in good agreement.

Finally, we note that our results on the scatter between halo mass and \textit{observed} stellar mass leave little room for \textit{intrinsic} scatter in stellar mass. For example, as discussed in  \cite{Bundy2015_ApJS_221_15}, the uncertainty in the observed stellar mass estimate of any of the three stellar mass estimates is of order $0.1 - 0.2 \, \mathrm{dex}$. On the other hand, based on theoretical models of galaxy formation, the intrinsic scatter in stellar mass at fixed halo mass is also expected to be at least $0.1 \, \mathrm{dex}$ \citep{Wechsler2018_ARAA_56_435}. A change in cosmological parameters or the inclusion of galaxy assembly bias in the modelling could potentially bring our results in better agreement with expectations.

\subsection{Precision of Stellar Mass Estimates}

In agreement with our findings, \cite{Tinker2017_ApJ_839_121} infer from BOSS CMASS galaxies that the spectroscopic Wisconsin stellar masses correlate more strongly with large-scale bias than the Granada stellar masses. We study new stellar mass estimates based on deeper photometry by the DESI Legacy Imaging Surveys that produce stellar mass-bias correlation similar to the Wisconsin mass estimates. As discussed in \cite{Tinker2017_ApJ_839_121}, the correlation of observed stellar mass with large-scale bias can be used to gauge the precision of different stellar mass estimates. The idea is that stellar mass and halo mass are strongly correlated, as is halo mass and large scale bias. Thus, the stellar mass estimate that produces the strongest clustering amplitude would correlate most strongly with halo bias, thereby halo mass and finally intrinsic stellar mass. We note that this argument would still work even in the presence of galaxy assembly bias, i.e. the correlation of intrinsic stellar mass with halo properties besides halo mass. However, the argument does not work if any secondary galaxy property at fixed intrinsic stellar mass correlates with large scale bias. For example, \cite{Berti2021_AJ_161_49} have shown that at fixed (observed) stellar mass, the specific star formation rate (sSFR) correlates with large-scale clustering, even if one only considers quiescent, red galaxies. Thus, for example, a bias of observed stellar mass as a function of sSFR at fixed intrinsic $M_\star$ could be an alternative explanation for the clustering differences between the three different stellar mass estimates.

\section{Conclusion}
\label{sec:conclusion}

In this analysis, we have provided new measurements of the scale and mass dependence of the lensing-is-low effect in the BOSS galaxy sample \citep{Leauthaud2017_MNRAS_467_3024, Lange2019_MNRAS_488_5771, Yuan2020_MNRAS_493_5551}. Our main result is that once cosmological parameters are fixed to those favoured by the \cite{PlanckCollaboration2020_AA_641_6} CMB analysis and a galaxy-halo model is fitted to the projected clustering of galaxies, the lensing is overpredicted by $\sim 35\%$, with no obvious dependence on halo mass in the range $\sim 10^{13.3} - 10^{13.9} \, h^{-1} \, M_\odot$ and or scale in the range $0.1 \, h^{-1} \, \mathrm{Mpc} < r_{\rm p} < 60 \, h^{-1} \mathrm{Mpc}$. These findings provide important constraints on possible solutions to the lensing-is-low phenomenon.

The lack of a strong halo mass dependence is qualitatively different than what is found in the recent DES Y1 cluster analysis \citep{Abbott2020_PhRvD_102_3509} but consistent with many plausible explanations for the lensing-is-low problem such as changes in cosmological parameters or baryonic feedback \citep{Lange2019_MNRAS_488_5771}. On the other hand, the lack of a strong scale dependence indicates that baryonic feedback or details of the galaxy-halo connection cannot fully explain the tension since those operate at scales below $r_{\rm p} \lesssim 5 \, h^{-1} \, \mathrm{Mpc}$. Additionally, as shown in \cite{Leauthaud2017_MNRAS_467_3024}, \cite{Lange2019_MNRAS_488_5771} and \cite{Yuan2020_MNRAS_493_5551}, both effects are unlikely to lower the lensing signal by $\sim 35\%$.

The apparent scale independence of the lensing-is-low effect over a broad range of scales provide tight constraints on possible new-physics explanations for the tension. In particular, models suppressing the matter power spectrum on comoving scales $k>0.05 h/$Mpc are unlikely to resolve the tension while preserving the observed scale independence of the effect. Complicating matters further is the fact that the comoving scales probed by the CMB have a significant overlap (see Fig.~\ref{fig:power_spectrum}) with those where the lensing-is-low effect occurs, implying that solutions modifying the amplitude of matter fluctuations at early times must be carefully vetted against CMB data. Solutions based on modifying the growth of matter fluctuations at late times must also explain the lack of scale dependence of the lensing-is-low effect while not running afoul of constraints on the late-time expansion history as probed by Type Ia supernovae and BAO. A detailed comparative analysis of possible solutions is left to future work.

Our study also provides valuable results on the galaxy-halo connection and stellar mass estimates. For example, we find that SDSS spectroscopic Wisconsin and DESI Legacy imaging photometric stellar mass estimates correlate more strongly with halo properties like halo mass and bias than the Granada mass estimates based purely on SDSS photometry. One possible explanation is that the former two mass estimates provide more precise estimates of the intrinsic stellar mass. Finally, for all three stellar mass estimates, we find evidence of mass segregation in the sense that more massive satellite galaxies orbit closer to the halo centre than less massive satellite galaxies.

In the future, we plan to combine the clustering and lensing measurements with estimates of the thermal and kinematic Sunyaev-Zeldovich effect around the same lenses. This can provide additional constraints on gas dynamics and the strength of baryonic feedback \citep{Amodeo2020_arXiv_2009_5558}. Similarly, we plan to investigate the lensing-is-low effect by cross-correlating BOSS lenses with other imaging surveys like DES, the Kilo Degree Survey (KiDS) or the Hyper Suprime Cam (HSC) survey. This will help to eliminate lensing systematics as a possible source of the unexpectedly low lensing signal. Preliminary results for BOSS LOWZ indicate that SDSS is accurate to within the quoted $6\%$ systematic error (Leauthaud et al., in prep.). Finally, combining galaxy-galaxy lensing signals with constraints from redshift-space clustering on non-linear scales is another promising avenue. First, redshift-space clustering could constrain the amount of galaxy assembly bias \citep{Lange2019_MNRAS_490_1870, Yuan2020_arXiv_2010_4182} and further reduce the uncertainty in the lensing predictions on non-linear scales. Additionally, redshift-space clustering is sensitive to the cosmological parameter combination $f \sigma_8$ where $f$ is the growth rate. Thus, combining redshift-space clustering with lensing could further break the $\Omega_{\rm m, 0} - \sigma_8$ compared to using only projected clustering and lensing, similar to the analysis of \cite{Troster2020_AA_633_10} on large scales.

\section*{Acknowledgements}

This work made use of the following software packages: {\sc matplotlib} \citep{Hunter2007_CSE_9_90}, {\sc SciPy}, {\sc NumPy} \citep{vanderWalt2011_CSE_13_22}, {\sc Astropy} \citep{AstropyCollaboration2013_AA_558_33}, {\sc Colossus} \citep{Diemer2015_ascl_soft_1016}, {\sc halotools} \citep{Hearin2017_AJ_154_190}, {\sc MultiNest} \citep{Feroz2008_MNRAS_384_449, Feroz2009_MNRAS_398_1601, Feroz2019_OJAp_2_10}, {\sc PyMultiNest} \citep{Buchner2014_AA_564_125}, {\sc Spyder} and {\sc GNOME \LaTeX}.

We acknowledge use of the lux supercomputer at UC Santa Cruz, funded by NSF MRI grant AST 1828315. This material is based on work supported by the U.D Department of Energy, Office of Science, Office of High Energy Physics under Award Number DE-SC0019301. AL acknowledges support from the David and Lucille Packard foundation, and from the Alfred P. Sloan foundation. HG acknowledges the support from the National Science Foundation of China (Nos. 11833005, 11922305). TLS is supported by by NSF Grant No.~2009377, NASA Grant No.~80NSSC18K0728, and the Research Corporation. The work of FYCR is supported by the NSF grant AST 2008696. RZ is supported by the Director, Office of Science, Office of High Energy Physics of the U.S. Department of Energy under Contract No. DE-AC02-05CH1123.

Funding for SDSS-III has been provided by the Alfred P. Sloan Foundation, the Participating Institutions, the National Science Foundation, and the U.S. Department of Energy Office of Science. The SDSS-III web site is http://www.sdss3.org/.

SDSS-III is managed by the Astrophysical Research Consortium for the Participating Institutions of the SDSS-III Collaboration including the University of Arizona, the Brazilian Participation Group, Brookhaven National Laboratory, Carnegie Mellon University, University of Florida, the French Participation Group, the German Participation Group, Harvard University, the Instituto de Astrofisica de Canarias, the Michigan State/Notre Dame/JINA Participation Group, Johns Hopkins University, Lawrence Berkeley National Laboratory, Max Planck Institute for Astrophysics, Max Planck Institute for Extraterrestrial Physics, New Mexico State University, New York University, Ohio State University, Pennsylvania State University, University of Portsmouth, Princeton University, the Spanish Participation Group, University of Tokyo, University of Utah, Vanderbilt University, University of Virginia, University of Washington, and Yale University.

The Legacy Surveys consist of three individual and complementary projects: the Dark Energy Camera Legacy Survey (DECaLS; NSF's OIR Lab Proposal ID \# 2014B-0404; PIs: David Schlegel and Arjun Dey), the Beijing-Arizona Sky Survey (BASS; NSF's OIR Lab Proposal ID \# 2015A-0801; PIs: Zhou Xu and Xiaohui Fan), and the Mayall z-band Legacy Survey (MzLS; NSF's OIR Lab Proposal ID \# 2016A-0453; PI: Arjun Dey). DECaLS, BASS and MzLS together include data obtained, respectively, at the Blanco telescope, Cerro Tololo Inter-American Observatory, The NSF's National Optical-Infrared Astronomy Research Laboratory (NSF's OIR Lab); the Bok telescope, Steward Observatory, University of Arizona; and the Mayall telescope, Kitt Peak National Observatory, NSF's OIR Lab. The Legacy Surveys project is honored to be permitted to conduct astronomical research on Iolkam Du'ag (Kitt Peak), a mountain with particular significance to the Tohono O'odham Nation.

The NSF's OIR Lab is operated by the Association of Universities for Research in Astronomy (AURA) under a cooperative agreement with the National Science Foundation.

This project used data obtained with the Dark Energy Camera (DECam), which was constructed by the Dark Energy Survey (DES) collaboration. Funding for the DES Projects has been provided by the U.S. Department of Energy, the U.S. National Science Foundation, the Ministry of Science and Education of Spain, the Science and Technology Facilities Council of the United Kingdom, the Higher Education Funding Council for England, the National Center for Supercomputing Applications at the University of Illinois at Urbana-Champaign, the Kavli Institute of Cosmological Physics at the University of Chicago, Center for Cosmology and Astro-Particle Physics at the Ohio State University, the Mitchell Institute for Fundamental Physics and Astronomy at Texas A\&M University, Financiadora de Estudos e Projetos, Fundacao Carlos Chagas Filho de Amparo, Financiadora de Estudos e Projetos, Fundacao Carlos Chagas Filho de Amparo a Pesquisa do Estado do Rio de Janeiro, Conselho Nacional de Desenvolvimento Cientifico e Tecnologico and the Ministerio da Ciencia, Tecnologia e Inovacao, the Deutsche Forschungsgemeinschaft and the Collaborating Institutions in the Dark Energy Survey. The Collaborating Institutions are Argonne National Laboratory, the University of California at Santa Cruz, the University of Cambridge, Centro de Investigaciones Energeticas, Medioambientales y Tecnologicas-Madrid, the University of Chicago, University College London, the DES-Brazil Consortium, the University of Edinburgh, the Eidgenossische Technische Hochschule (ETH) Zurich, Fermi National Accelerator Laboratory, the University of Illinois at Urbana-Champaign, the Institut de Ciencies de l'Espai (IEEC/CSIC), the Institut de Fisica d'Altes Energies, Lawrence Berkeley National Laboratory, the Ludwig-Maximilians Universitat Munchen and the associated Excellence Cluster Universe, the University of Michigan, the National Optical Astronomy Observatory, the University of Nottingham, the Ohio State University, the University of Pennsylvania, the University of Portsmouth, SLAC National Accelerator Laboratory, Stanford University, the University of Sussex, and Texas A\&M University.

BASS is a key project of the Telescope Access Program (TAP), which has been funded by the National Astronomical Observatories of China, the Chinese Academy of Sciences (the Strategic Priority Research Program ''The Emergence of Cosmological Structures`` Grant \# XDB09000000), and the Special Fund for Astronomy from the Ministry of Finance. The BASS is also supported by the External Cooperation Program of Chinese Academy of Sciences (Grant \#114A11KYSB20160057), and Chinese National Natural Science Foundation (Grant \# 11433005).

The Legacy Survey team makes use of data products from the Near-Earth Object Wide-field Infrared Survey Explorer (NEOWISE), which is a project of the Jet Propulsion Laboratory/California Institute of Technology. NEOWISE is funded by the National Aeronautics and Space Administration.

The Legacy Surveys imaging of the DESI footprint is supported by the Director, Office of Science, Office of High Energy Physics of the U.S. Department of Energy under Contract No. DE-AC02-05CH1123, by the National Energy Research Scientific Computing Center, a DOE Office of Science User Facility under the same contract; and by the U.S. National Science Foundation, Division of Astronomical Sciences under Contract No. AST-0950945 to NOAO.

\section*{Data Availability}

The Abacus simulations used in this article are available in the Abacus Cosmos database at  \url{https://lgarrison.github.io/AbacusCosmos/}. The SDSS data analysed is available at \url{https://data.sdss.org/sas/dr12/boss/lss/} and the data of the DESI Legacy Imaging Surveys at \url{https://www.legacysurvey.org/}. Upon publication, the clustering and lensing measurements analysed in this work will be publicly available at \url{https://johannesulf.github.io/data/}. All derived data generated in this research as well as code used will be shared on reasonable request to the corresponding author.

\bibliographystyle{mnras}
\bibliography{bibliography}

\label{lastpage}

\end{document}